\documentclass[11pt]{article}

\usepackage{amsfonts}
\usepackage{latexsym}
\usepackage{amssymb}
\usepackage{amscd}
\usepackage{amsmath}

\newlength{\dinwidth}
\newlength{\dinmargin}
\def\be{\begin{equation}}
\def\ee{\end{equation}}
\def\ben{\begin{displaymath}}
\def\een{\end{displaymath}}
\def\baa{\begin{eqnarray}}
\def\eaa{\end{eqnarray}}

\def\ba{\begin{array}}
\def\ea{\end{array}}

\makeatletter
\@addtoreset{equation}{section}
\makeatother

\def\phi{\varphi}

\def\a{\alpha}
\def\b{\beta}

\def\de{{\delta}}
\def\e{\epsilon}
\def\l{\lambda}

\def\Th{\Theta}

\def\Acal{{\cal A}}
 
\def\Ucal{{\cal U}}
\def\Lhat{{\hat{{\cal L}}}}

\def\Pcal{{\cal P}}
\def\Ucal{{\cal U}}
\def\Ecal{{\cal E}}
\def\Wcal{{\cal W}}

\def\be{\begin{equation}}
\def\ee{\end{equation}}
\def\f{\frac}
\def\la{\label}

\def\CP1{{\mathbb C}P^1}

\def\la{\label}

\def\f{\frac}
\def\L{{\cal L}}
\def\p{\partial}

\def\tr{{\rm tr}}
\def\log{\ln}

\def\la{\label}

\def\f{\frac}
\def\p{\partial}
\def\res{{\rm res}}

\def\det{{\rm det}}

\def\L{{\cal L}}
\def\B{{\bf B}}

\def\U{{V_1}}
\def\V{{V_2}}

\def\x{{f}}
\def\y{{g}}

\def\xp{{P}}
\def\yp{{Q}}

\def\ix{{\infty_f}}
\def\iy{{\infty_g}}

\def\dU{{d_1}}
\def\dV{{d_2}}

\def\Y1{{Y^{(1)}}}
\def\pib{f}

\def\M{{\cal M}}
\def\B{{\bf B}}

\def\lp{{x}}
\def\lpy{{y}}

\def\blangle{{\Big\langle}}
\def\brangle{{\Big\rangle}}

\def\xt{{\tilde{x}}}

\newtheorem{remark}{Remark}
\newtheorem{theorem}{Theorem}

\newtheorem{lemma}{Lemma}
\newtheorem{proposition}{Proposition}

\begin{document}

\begin{center}
\hskip9.0cm {\large Preprint SPhT-T04/020}
\vskip1.0cm
{\LARGE Genus one contribution to free energy in hermitian two-matrix model}
\vskip1.0cm
{\large B.Eynard, A.Kokotov, D.Korotkin}
\vskip0.5cm
$^1$ Service de Physique The\'eorique, CEA/Saclay, Orme des Merisier
F-91191 Gif-sur-Yvette Cedex, France\\
$^2$ Concordia University, 7141 Sherbrooke West, Montreal H4B1R6,
Montrteal, Quebec, Canada
\end{center}
\vskip1.0cm

{\bf Abstract.} We compute an the  genus 1 correction to free energy of
Hermitian two-matrix model in terms of theta-functions
associated to spectral curve arising in large N limit. We discuss the
relationship of this expression to
isomonodromic tau-function, Bergmann tau-function on
Hurwitz spaces, G-function of Frobenius manifolds and determinant of
Laplacian in a singular metric  over spectral curve.

\section{Two-matrix models: introduction}

In this paper we study the partition function of multi-cut two-matrix model:
\be
Z_N\equiv e^{-N^2F}:= \int dM_1 dM_2 e^{-N\tr\{V_1(M_1)+V_2(M_2)-M_1 M_2\}}
\la{part}
\ee
where the integral is taken over all independent entries of two hermitian matrices $M_1$ and $M_2$ 
such that the eigenvalues of $M_1$ are concentrated over a finite set of intervals (cuts) with 
given filling fractions.
This integral is to be understood as a formal asymptotic series in $N$ and
in the coefficients of the two potentials $V_1$ and $V_2$.
As a formal series, the questions of convergence of the matrix
integral is irrelevant, and the model can be extended to matrices with
eigenvalues constrained on contours in the complex plane.

Such asymptotic series play an important role in physics, as generating functions of statistical physics on random discretized polygonal surfaces, 
i.e. a simplified model of euclidean 2D quantum gravity \cite{DGZ, KazakovIsing}.
The large $N$ expansion $F=\sum_{G=0}^{\infty} N^{-2G} F^{G}$ 
($N$  is the matrix size), called topological expansion,
is one of the cornerstones of the theory, since $F^{G}$ has the
meaning of generating function for random discretized polygonal surfaces of genus $G$.
Double scaling limits of these models correspond to statistical physics models on continuous surfaces, with conformal invariance properties.
Matrix models thus provide realizations of minimal $(p,q)$ conformal models.
The 1-matrix model was shown to correspond to pure gravity (i.e. $q=2$), and
the 2-matrix model was introduced as it produces all $(p,q)$ minimal models.

Recently, the interest in large $N$ matrix models
was renewed as it was understood \cite{DV}, that the large
$N$ free energy of matrix models is the low energy effective action
for some string theories.
The computation of $1/N^2$ expansion for both one-matrix and two-matrix models is based on the loop equations,
which was first derived for 1-matrix 1-cut in \cite{Chekhov1}, then
for 1-matrix 2-cuts in \cite{Akemann,KMT}, and recently derived in \cite{Kostov,Chekhov,DST} for 1-matrix model multicut, and in \cite{Eynard1,Eynard2} for two-matrix model 1 and 2-cuts. 
Here, we will extend the results of \cite{Eynard1,Eynard2} for an
arbitrary number of cuts i.e. for an arbitrary (up to maximal) genus
of the spectral curve.

Writing down polynomials $\U$ and $\V$ in the form
\be
\U(x)=\sum_{k=1}^{d_1+1} \f{u_k}{k} x^k\;\;,\hskip0.6cm 
\V(y)=\sum_{k=1}^{d_2+1} \f{v_k}{k} y^k\;,
\ee
we shall use the following standard  notations for operators of differentiation with respect
to their coefficients:
\be
\f{\de}{\de\U(x)}\Big|_x:=\sum_{k=1}^{d_1+1}x^{-k-1} k\p_{u_k}\;\;,\hskip0.8cm
\f{\de}{\de\V(y)}\Big|_y:=\sum_{k=1}^{d_2+1}y^{-k-1} k\p_{v_k}\;.
\ee
These notations will be used below to shorten some of the formulas;  by definition
the equality
\be
\f{\de F}{\de\U(x)}\Big|_x = H(x)\hskip0.5cm
{\rm means\hskip0.3cm that}                \hskip0.5cm
\f{\p F}{\p u_k}= \f{1}{2\pi i k}\oint_{x=\infty}x^k H(x)dx\;,\hskip0.5cm
k=1,\dots,d_1+1\;;
\la{ex2}
\ee
a detailed discussion of this notation is contained in \cite{Marco2}.
In fact, formally it is much more convenient not to cut the functions $\U$ and $\V$ 
to polynomials, but instead consider the Laurent series
\be
\U(x)=\sum_{k=1}^{\infty} \f{u_k}{k} x^k\;\;,\hskip0.6cm 
\V(y)=\sum_{k=1}^{\infty} \f{v_k}{k} y^k\;.
\la{UVinf}
\ee 
In this case we  have the formal relations
\be
\f{\de \U(x)}{\de \U(\xt)}=\f{1}{\xt-x}\;,\hskip0.6cm 
\f{\de \U'(x)}{\de \U(\xt)}=\f{1}{(\xt-x)^2}\;,
\la{deltaf}
\ee
which are implicitly used in the derivation of loop equation. However, the convergency problem with
considering all coefficients in the infinite sums (\ref{UVinf}) to be independent variables
forces us to understand all relations involving the operators $\de/\de\U(x)$ and $\de/\de\V(y)$ 
in the sense of (\ref{ex2}).

Consider the resolvents (also understood as formal power series)
\be
\Wcal(x)=\f{1}{N}\blangle \tr\f{1}{x-M_1}\brangle
\hskip0.7cm {\rm and}\hskip0.7cm
\tilde{\Wcal}(y)=\f{1}{N}\blangle \tr\f{1}{y-M_2}\brangle\;.
\la{resolv}
\ee
As a corollary of (\ref{deltaf}), 
the free energy of two-matrix model (\ref{part}) satisfies the following equations with respect to
coefficients of polynomial $\U$:
\be
\f{\delta F}{\delta \U(x)} =\Wcal(x)\;,\hskip0.7cm
\f{\delta F}{\delta \V(y)} =\tilde{\Wcal}(y)\;,
\la{varint}
\ee
 valid in the sense of (\ref{ex2}).

Assuming existence of $1/N^2$ expansion, the equations (\ref{varint}) 
were solved in \cite{Marco1} in the zeroth order  in terms of 
holomorphic objects associated to the ``spectral curve'' which arises in $N\to \infty$ limit.
The next coefficient $F^1$ was computed in \cite{Eynard1} if the genus of spectral curve equals zero,
and in \cite{Eynard2} if the genus equals one.
The main result of this paper is an expression for $F^1$ for an arbitrary genus of ``spectral curve'',
which we find using loop equations.  We compute $F^1$ using the
algebro-geometric framework (the spectral curve and corresponding machinery) 
which arises already in zeroth order approximation.

The spectral curve is defined by the following equation:
\be
\Ecal^{0}(x,y):= (\U'(x)-y)(\V'(y)-x) -\Pcal^{0}(x,y)+1=0\;,
\la{Lint}
\ee
 where polynomial of two variables 
$\Pcal^{0}(x,y)$ is the zeroth order term in $1/N^2$ expansion of polynomial  
\be
\Pcal(x,y):=\f{1}{N}\blangle \tr\f{\U'(x)-\U'(M_1)}{x-M_1}\f{\V'(y)-\V'(M_2)}{y-M_2}\brangle\;;
\ee
the point $P$ of this curve is the pair of complex numbers $(x,y)$ satisfying (\ref{Lint}).

The spectral curve (\ref{Lint}) comes together with two meromorphic functions
$\x(P)=x$ and $\y(P)=y$, which project it down to $x$ and $y$-planes, respectively.
These functions have poles only at two points of $\L$, called $\ix$ and $\iy$: at $\ix$ function $\x(P)$ has
simple pole, and function $\y(P)$ - pole of order $d_1$ with singular part equal to $\U'(\x(P))$. At
$\iy$ the function $\y(P)$ has simple pole, and function $\x(P)$ - pole  of order $d_2$ with singular part
equal to $\V'(\y(P))$. Therefore, one gets  the moduli space ${\cal M}$ of triples $(\L,\x,\y)$, where functions $\x$ and $\y$ have this pole structure. The natural coordinates on this moduli space are coefficients of polynomials $\U$ and $\V$ and $g$ numbers,
called ``filling fractions" $\e_\a=\f{1}{2\pi i}\oint_{a_\a}\y d\x$, where $a_\a$ are (chosen in some way)
canonical cycles on $\L$.  



Denote the zeros of differential $d\x$ by $\xp_1,\dots,\xp_{m_1}$ ($m_1=d_2+2g+1$) (these points play the 
role of ramification points if we realize $\L$ as branched covering by function $\x(P)$); their projections on 
$\x$-plane are the branch points, which we denote we denote by $\l_j:=\x(\xp_j)$ .
  The zeros of the differential $d\y$ (the ramification points if we consider $\L$ 
as covering defined by function $\y(P)$) we denote by 
$\yp_1,\dots,\yp_{m_2}$  ($m_2=d_1+2g+1$); 
their projections on $\y$-plane (the branch points) we denote by $\mu_j:=\y(\yp_j)$.
We shall assume that our potentials $\U$ and $\V$ are generic i.e. all zeros of differentials $d\x$ and $d\y$ are simple and distinct.

If is well-known \cite{Marco1} how to express all standard
algebro-geometrical objects on $\L$ in terms of the previous  data. In particular,
the Bergmann bidifferential $B(P,Q)=d_P d_Q\log E(P,Q)$ ($E(P,Q)$ is the prime-form), 
can be 
represented as follows: 
\be
 B(P,Q)=\f{\de \y(P)}{\de \U(\x(Q))}\Big|_{\x(Q)} d\x(P) d\x(Q)
\la{Bx}
\ee
(see \cite{Marco1} for the proof).
The Bergmann bidifferential has the following 
behaviour near diagonal $P\to Q$:
\be
B(P,Q)=\left\{\f{1}{(\tau(P)-\tau(Q))^2}+ \f{1}{6}S_B(P)+ o(1)\right\} d\tau(P) d\tau(Q)\;,
\la{Bdia}
\ee
where $\tau(P)$ is some local coordinate; $S_B(P)$ is the Bergmann projective connection 
($S_B(P)$ transforms as quadratic differential under M\"obius transformations; under an arbitrary coordinate transformation
an appropriate Schwarzian derivative is added to it).

Consider also the  four-differential $D(P,Q)=d_P d_Q^3\log E(P,Q)$, which has on the diagonal 
the pole of 4th degree:
$D(P,Q)=\{ 6(\tau(P)-\tau(Q))^{-4}+ O(1)\} d\tau(P) (d\tau(Q))^3$.
From $B(P,Q)$ and $D(P,Q)$ it is easy to construct meromorphic normalized (all $a$-periods vanish) 
$1$-forms on $\L$ 
with single pole; in particular, if the pole coincides with ramification point $\xp_k$, the natural 
local parameter
near $\xp_k$ is $\lp_k(P)=\sqrt{\x(P)-\l_k}$, and the following objects:
\be
B(P,\xp_k):=\f{B(P,Q)}{d\lp_k(Q)}\Big|_{Q=P_k}\;,\hskip0.7cm
D(P,\xp_k):=\f{D(P,Q)}{(d\lp_k(Q))^3}\Big|_{Q=P_k}
\la{BDkdef}
\ee
are meromorphic normalized 1-forms on $\L$ with single pole at $\xp_k$ and the following singular 
parts:
\be
B(P,\xp_k)=\left\{\f{1}{\lp_k(P)^2}+ \f{1}{6}S_B(P_k)+ o(1)\right\} d\lp_k(P)\;;\hskip0.7cm
D(P,\xp_k)=\left\{\f{6}{\lp_k(P)^4}+ O(1)\right\} d\lp_k(P)
\la{BDk}
\ee
as $P\to \xp_k$, where $S_B(P_k)$  is the Bergmann projective connection computed at the branch 
point $P_k$
with respect to the local parameter $\lp_k(P)$.

Equations (\ref{varint}) in order $1/N^2$ look as follows (we write only equations with respect to $\U$):
\be
\f{\de F^{1}}{\de\U(\x(P))}=-Y^{1}(P)\;,
\la{F1Yint}
\ee
where the $Y^{1}$ is the $1/N^2$ contribution to the resolvent ${{\cal W}}$. The function $Y^{1}$
can be computed using the
loop equations \cite{Eynard1}, which leads to the  following expression:
\be
\Y1(P)d\x(P) =\sum_{k=1}^{m_1}\left\{-\f{1}{96\y'(\xp_k)}D(P,\xp_k) + \left[\f{\y'''(\xp_k)}{96\y'^2(\xp_k)} -\f{S_B(\xp_k)}{24\y'(\xp_k)}\right] B(P,\xp_k)\right\}\;.
\la{Y1int}
\ee
The solution of (\ref{F1Yint}), (\ref{Y1int}) invariant with respect to the projection change
(i.e. which satisfies also the required equations with respect to $\V$), looks as follows:
\be
F^{1}=\f{1}{24}\log\left\{\tau^{12}_{\x} (v_{\dV+1})^{1-\f{1}{\dV}}\prod_{k=1}^{m_1} d\y(\xp_k) 
\right\}+\f{d_2+3}{24}\log d_2\;,
\la{F1int}
\ee
where $\tau_\x$ is the so-called Bergmann tau-function on Hurwitz space, which satisfies the 
following system of equations with respect to the branch points $\l_k$:
\be
\f{\p}{\p \l_k}\log\tau_{\x} = -\f{1}{12} S_B(P_k)\;.
\la{bergint}
\ee
The Bergmann tau-function (\ref{bergint}) appears in many important problems: it coincides with 
isomonodromic
tau-function of Hurwitz Frobenius manifolds \cite{Dub92}, and gives the main contribution to 
$G$-function
(solution of Getzler equation)
of these Frobenius manifolds; it gives the most non-trivial term in isomonodromic tau-function of 
Riemann-Hilbert problem with quasi-permutation monodromies. Finally, its modulus square essentially 
coincides with determinants of Laplace
operator in metrics with conical singularities over Riemann surfaces \cite{KKprep}. 
The solution of the system (\ref{bergint}) was found in \cite{Duke} and can be described as follows.

Define the divisor $(d\x)=-2\ix-(d_2+1)\iy+\sum_{k=1}^{m_1} P_k:=\sum_{k=1}^{m_1+2}r_k D_k$.
Choose some initial point 
$P\in\Lhat$ and introduce corresponding vector of Riemann constants $K^P$ and Abel map 
$\Acal_\a(Q)=\int_P^Q w_\a$
($w_\a$ form the basis of normalized holomorphic 1-forms on $\L$).
Since all zeros of differential $d\x$ have multiplicity 1, we can always choose the fundamental 
cell $\Lhat$
of the universal covering of $\L$  in such a way 
that $\Acal((d\x))=-2K^P$ (for an arbitrary choice of fundamental domain these two vectors 
coincide only up to
an integer combination of periods of holomorphic differentials), where the Abel map is computed 
along the path which does not intersect the boundary of $\Lhat$.

The main ingredient of the Bergmann tau-function is the following
holomorphic multivalued $(1-g)g/2$-differential ${{\cal C}}(P)$ on $\L$:
\be
{{\cal C}}(P):=\f{1}{W(P)}{\sum_{\a_1,\dots,\a_g=1}^g 
\frac{\partial^g\Theta(K^P)}{\p z_{\a_1}\dots \p z_{\a_g}} w_{\a_1}(P)\dots w_{\a_g}(P)}\;.
\ee
where
\be
W(P):= {\rm \det}_{1\leq \a, \b\leq g}||w_\b^{(\a-1)}(P)||
\la{Wronks1}
\ee
denotes the Wronskian determinant of holomorphic differentials
 at point $P$. Introduce also the quantity ${\cal Q}$  defined by
\begin{equation}
{\cal Q} = [d\x(P)]^{\f{g-1}{2}} {{\cal C}}(P)
\prod_{k=1}^{m+2}[ E (P,D_k)]^{\f{(1-g)r_k}{2}}\;,
\la{Fdef1}
\end{equation}
which is independent of the point $P\in \L$. 
Then the Bergmann tau-function (\ref{bergint}) of Hurwitz space  is given by the following expression:
\begin{equation}
\tau_{\x} = {{\cal Q}}^{2/3}  \prod_{k,l=1\;\;k< l}^{m+n} [E(D_k, D_l)]^{\frac{r_k r_l}{6}} \;; 
\la{taui}
\end{equation}
together with (\ref{F1int}) this gives the answer for $1/N^2$ correction in two-matrix model. 

If potential $\V$ is quadratic, integration with respect to $M_2$ in (\ref{part}) can be taken explicitly, and
the free energy (\ref{F1int}) gives rise to the free energy of one-matrix model. The spectral curve $\L$ in this
case  becomes hyperelliptic, and the formula (\ref{F1int}) gives, using the expression for $\tau_\x$ obtained in 
\cite{KitKor}:
\be
F^1=\f{1}{24}\log\left\{\Delta^3\,(\det{\bf A})^{12}  \prod_{k=1}^{2g+2} \y'(\l_k)\right\}\;,
\la{F1hypint}
\ee
where $\l_k$, $k=1,\dots,2g+2$ are branch points of $\L$; $\Delta$ is their Wronskian determinant; ${\bf A}$ 
is the matrix of $a$-periods of non-normalized holomorphic differentials on $\L$.

The paper is organised as follows. In section 2, following \cite{Eynard1},
 we write down the loop equations for two-matrix model, and discuss the spectral curve and associated objects
which arise in the zeroth order in $1/N^2$ expansion. Here we derive also some new variational
formulas, which will be used later in computation of $1/N^2$ correction to free energy.
In section 3 we solve the loop equations in  $1/N^2$ approximation. 
Here we also express $F^1$ in terms of Bergmann tau-function on Hurwitz spaces introduced in \cite{Dub92,Koro00}.
In section 4 we recall the explicit expression for Bergmann tau-function \cite{Duke}, and find its transformation law
under the change of projection of the spectral curve to $\CP1$.
This allows to get the formula for $F^1$ which satisfies the full set of variational equations with respect to 
polynomials $\U$ and $\V$.
In section 5 we derive variational equation of $F^1$ with respect to filling fractions.
In section 6 we discuss the links between $F^1$ and other related objects: determinant of Laplace operator,
$G$-function of Frobenius manifolds and isomonodromic tau-function of fuchsian system with quasi-permutation monodromies.

\section{Loop equations: leading term}

Introduce the function
\be
Y(x)= V_1'(x)-\Wcal(x)
\la{Y}
\ee
In terms of function $Y$ equations (\ref{varint}) for free energy can be written as follows: 
\be
\f{\delta F}{\delta \U(x)} =\U'(x)-Y(x)\;,
\la{varfU}
\ee
as well as (\ref{varint}),  valid in the sense of (\ref{ex2}).

To make use of variational formula (\ref{varfU}) we need to get some information about  the function $Y(x)$.
This information is in principle contained in the 
loop equations, which  follow from  reparametrization invariance of the partition function
(\ref{part})  (see \cite{Eynard1} for details).
To write them down, apart from resolvent $\Wcal(x)$ (\ref{resolv}), we need to introduce the following objects:

\begin{itemize}
\item Polynomial  $\Pcal(x,y)$:
\be
\Pcal(x,y):=\f{1}{N}\blangle \tr\f{\U(x)-\U(M_1)}{x-M_1}\f{\V(y)-\V(M_2)}{y-M_2}\brangle
\la{Pxy}
\ee

\item Polynomial  $\Ecal(x,y)$
\be
\Ecal(x,y):=(\U(x)-y)(\V(y)-x)-\Pcal(x,y)+1
\la{Exy}
\ee

\item Function $\Ucal(x,y)$, which is a polynomial in $y$:
\be
\Ucal(x,y):=\f{1}{N}\blangle \tr\f{1}{x-M_1}\f{\V'(y)-\V'(M_2)}{y-M_2}\brangle
\la{Uxy}
\ee

\item Function $\Ucal(x,y,z)$, which is also  a polynomial in $y$:
\be
\Ucal(x,y,z):=\f{\de \Ucal(x,y)}{\de \U(z)}= 
\blangle \tr\f{1}{x-M_1}\f{\V'(y)-\V'(M_2)}{y-M_2}\tr\f{1}{z-M_1}\brangle
-N^2 \Ucal(x,y) \Wcal (z)
\la{Uxyxt}
\ee
\end{itemize}

Now we are in position to write down the  loop equation  
\be
\Ucal(x,y)=x-\V'(y)+\f{\Ecal(x,y)}{y-Y(x)}-\f{1}{N^2}\f{\Ucal(x,y,x)}{y-Y(x)}
\la{loop1}
\ee
which arises as a corollary of reparametrization invariance of the matrix integral
(\ref{part}) \cite{Eynard1}.

The residue at $y=Y(x)$ of (\ref{loop1}) leads to the following loop equation
(for polynomials of degree 3 this equation was first derived in \cite{Staudacher}) for function 
$Y(x):=\U'(x)-\Wcal(x)$:
\be
\Ecal^0(x,Y(x))=\f{1}{N^2} \Ucal(x,Y(x),x)\;.
\la{master}
\ee
 To use the loop equation effectively we need to consider the $1/N^2$ expansion of all of their ingredients.

\subsection{Leading order  term: algebro-geometric framework }

Assume that the function $Y$ admits an expansion into a power series in $1/N^2$:
\be
Y(x)=Y^{0}+\f{1}{N^2} Y^{1}+\dots\;.
\la{YY0}
\ee
Then in the leading order the master loop equation (\ref{master}) turns into algebraic equation 
in two variables: $x$ and $Y^{(0)}(x)$:
$$
\Ecal(x,Y^{0}(x))=0\;,
$$
where 
\be
\Ecal^{0}(x,y)= (\U'(x)-y)(\V'(y)-x) -\Pcal^{0}(x,y)+1\;.
\la{E0}
\ee
The polynomial equation 
\be
\Ecal^{0}(x,y)=0 
\la{alcur}
\ee
defines an algebraic curve $\L$ of some genus $g$, which we call
``spectral curve'' (if the spectral curve is non-singular, it has
``maximal genus'' equal to $d_1 d_2-1$);
the point $P$ of this curve is a pair of complex numbers $(x,y)$
satisfying the polynomial equation (\ref{alcur}).
Therefore,
 $Y^{0}$ can be considered as multi-valued function of $x$.
The curve $\L$ comes together with two meromorphic functions on it:
function $\x(P)=x$ and function $\y(P)=y$ ($\equiv Y^0(x)$). 
Since polynomial $\Pcal$ (\ref{Pxy})  and function $\Ecal$ (\ref{Exy}) are symmetric with respect to substitution
$x\leftrightarrow y$, $\U\leftrightarrow \V$, the same algebraic curve
appears if we  write down the loop equations
for $X(y):=\V'(y)-\f{\delta F}{\delta \V(y)}$.

Analytical properties of functions $\x(P)$ and $\y(P)$ on $\L$ are well-known (see \cite{Marco1,Marco2} and
references therein).
Namely, $\x(P)$ and $\y(P)$ are meromorphic functions on $\L$ having poles only 
at two marked points $\ix$ and $\iy$
with the following pole structure: function $\x(P)$ has simple pole at $\ix$  and pole of order $d_1$
at $\iy$; function $\y(P)$ has simple pole at $\iy$ and pole of order $d_2$ at $\ix$. Therefore, near $\ix$ we can
write the singular part of $\y(P)$ as polynomial of $\x(P)$; near $\iy$ we can represent  
 the singular part of $\x(P)$ as polynomial of $\y(P)$; coefficients of these polynomials are given by $\U'$ and
$\V'$, respectively:
\be
\y(P)=\U'(\x(P)) -\f{1}{\x(P)}+ O(\x^{-2}(P))\hskip0.5cm {\rm as}\hskip0.5cm P\to\ix\;,
\la{assix}
\ee
\be
\x(P)=\V'(\y(P)) -\f{1}{\y(P)}+ O(\y^{-2}(P))\hskip0.5cm {\rm as}\hskip0.5cm P\to\iy\;.
\la{assiy}
\ee

The dimension of the moduli space of triples $(\L,f,g)$  satisfying these conditions equals $d_1+d_2+g+2$.
Let us choose on $\L$ a canonical basis of cycles $(a_\a,b_\a)$.
Then coordinates on the space $\M$ can be chosen as follows:

\begin{itemize}
\item $d_1+1$ coefficients $u_1,\dots, u_{d_1+1}$ of polynomial $\U'$. 

\item $d_2+1$ coefficients $v_1,\dots, v_{d_2+1}$ of polynomial $\V'$.


\item The ``filling fractions" 
\be
\e_\a := \f{1}{2\pi i}\oint_{a_\a} \y d\x\;.
\la{filfr}
\ee
\end{itemize}

In strictly physical situation  potentials $\U$ and $\V$ should be such that,
considering $\L$ as a covering defined by function $f$,  one can single out the ``physical'' sheet (which includes point $\ix$)
such that all $a$-cycles lie on this sheet and each $a$-cycle encircles exactly one branch cut 
(all corresponding branch points must be real if potentials $\U$ and $\V$ are real). Similar requirement comes from
$\y$-projection of $\L$.
However, here we don't impose these ``physical'' requirements i.e.  consider the ``analytical continuation''
of physical sector, in the spirit of \cite{DV}.

Nevertheless, the sheet of the curve $\L$ (realized as $d_2+1$-sheeted
branched covering by function $f$), which contains the point $\ix$, is
called the ``physical" sheet; the physical sheet is well-defined at least in some neighbourhood of
$\ix$. Fixing some splitting of $\L$ into $d_2+1$ sheets, we   denote by $x^{(k)}$ 
($k=1,\dots,d_2+1$) the point of $\L$ belonging to
$k$th sheet such that $\x(x^{(k)})=x$; we assume that point $x^{(1)}$ belongs to the physical sheet of $\L$ i.e.
$x^{(1)}\to\ix$ as $x\to\infty$.

The polynomial $\Ecal^0(x,y)$ defining the spectral curve $\L$ (\ref{alcur})  can also be rewritten as follows:
\be
\Ecal^0(x,y)=-v_{d_2+1}\prod_{k=1}^{d_2+1}(y-g(x^{(k)}))
\la{Ezeros}
\ee
The proof of (\ref{Ezeros}) is simple: function $\Ecal^0$ is given by (\ref{E0}); since $\Pcal^0$ is a polynomial
of degree $d_2-1$ with respect to $y$, function $\Ecal^0$ is a polynomial of degree $d_2+1$ in $y$; its zeros
are $Y^0(x^{(k)})$ by definition of points $x^{(k)}$.
Comparison of  coefficient in front of $y^{d_2+1}$ leads to (\ref{Ezeros}).


\subsection{Some variational formulas}

If a Riemann surface is realized as a branched covering of Riemann sphere, the branch points can be
used as natural parameters on the moduli space, and it is easy to differentiate all objects introduced above with
respect to the branch points. The answer is given by Rauch variational formulas (\cite{Rauch}, for a simple proof
see \cite{MPAG}).
However, on our moduli space the set of natural coordinates is given
by coefficients of polynomials $V_1$ and $V_2$ and filling
fractions. To differentiate all interesting objects with respect to
these coordinated we need to know
the matrix of derivatives of branch points (we shall consider only $\{\l_k\}$) with respect to coefficients
of $\U$, $\V$ and filling fractions. This matrix was computed in \cite{Marco1}; below we rederive some of these
formulas, and prove new variational formulas, required in our context.


In \cite{Marco1} equations (\ref{varfU}), together
with their counterpart with respect to $\V(y)$,  were solved in
the leading term i.e. it was found the solution of the system
$$
\f{\de F^{0}}{\de \U(\x(P))}\Big|_{\x(P)} =\U'(\x(P))-\y(P)
$$
$$
\f{\de F^{0}}{\de \V(\y(P))}\Big|_{\y(P)} =\V'(\y(P))-\x(P)
$$
which {\it a posteriori} turns out to satisfy also the following equations with respect to filling fractions:
$$
\f{\p F^{0}}{\p \e_\a}=\Gamma_\a := \oint_{b_\a}\y(P)d\x (P)\;.
$$

To find solution of  the equations  (\ref{varfU}) in order $1/N^2$, together with their counterpart
with respect to $\V(y)$)
 we shall need 
\begin{lemma}\la{deryU}
The following variational formulas take place:
\be
-\f{\de\l_k}{\de \U(\x(P))} \y'(P_k) d\x(P)= B(P,P_k)\;,
\la{zerothor}
\ee
\be
\f{\de\{\y'(\xp_k)\}}{\de \U(\x(P))}\Big|_{\x(P)} d\x(P)=\f{1}{4}\left\{D(P,\xp_k)-\f{\y'''(\xp_k)}{\y'(\xp_k)} B(P,\xp_k)\right\}
\la{yV1}
\ee
\end{lemma}
{\it Proof.} 
We start from formula  (\ref{Bx}) for the Bergmann bidifferential:
\be
 B(P,Q)=\f{\de \y(P)}{\de \U(\x(Q))}\Big|_{\x(Q)} d\x(P) d\x(Q)\;.
\la{Bx1}
\ee
We want to rewrite this formula in the limit $Q\to P_k$ using the local parameter
$\lp_k(Q)=\sqrt{\x(Q)-\l_k}$. As  the first step we notice that for any coordinate $t$ on our moduli space
we have the following identity:
\be
\y_t(Q)|_\x(Q) d\x(Q)=\y_t(Q)|_{\lp_k(Q)}d\x(Q)-\x_t(Q)|_{\lp_k(Q)}d\y(Q)\;,
\la{thermod}
\ee
which follows from differentiation of composite function $\y(t,\x(\lp_k,t))$ with respect to $t$ using the chain rule.
In particular, 
\be
\f{\de\y(Q)}{\de \U(\x(P))}\Big|_\x(Q) d\x(Q)=\f{\de\y(Q)}{\de \U(\x(P))}\Big|_{\lp_k(Q)}d\x(Q)-
\f{\de\y(Q)}{\de \U(\x(P))}\Big|_{\lp_k(Q)}d\y(Q)\;.
\la{thermod1}
\ee
Consider now first several terms of local expansion of $\y(Q)$, $d\y(Q)$ and $B(P,Q)$ as $Q\to P_k$
(prime denotes derivative with respect to $\lp_k :=\lp_k(Q)$):
\be
\y(Q)= \y(\xp_k) + \y'(\xp_k) \lp_k +
\dots\;,
\la{yexp}
\ee
\be
d\y(Q)=\{\y'(\xp_k) + \y''(\xp_k)\lp_k +\f{1}{2}\y'''(\xp_k)\lp_k^2+\dots\} d\lp_k\;,
\la{dyexp}
\ee
\be
B(P,Q)=\{B(P,P_k)+B'(P,P_k)\lp_k+\f{1}{2}B''(P,P_k)\lp_k^2+\dots\}d\lp_k\;.
\ee
Taking into account that $\x(Q)=\lp^2_k+\l_k$, and substituting these relations into (\ref{thermod1}),
we get in the  order zero the formula (\ref{zerothor}).

The first order terms give relation which defines the dependence of $\y(P_k)$ on $\{u_k\}$:
\be
\left\{2 \f{\de g(P_k)}{\de \U(\x(P))}  -\f{\de\l_k}{\de \U(\x(P))} \y''(P_k)\right\}
d\x(P)=B'(P,P_k)\;;
\ee
we present this relation only for completeness, since it  will not be used below.

Finally,
collecting the coefficients in front of $\lp_k^2$, we get
$$
2 \f{\de \y'(P_k)}{\de \U(\x(P))}  -\f{1}{2}\f{\de \l_k}{\de \U(\x(P))}  \y'''(P_k)=\f{1}{2}\f{B''(P,P_k)}{d\x(P)}\;,
$$
which leads to (\ref{yV1}) after using (\ref{zerothor}).


\section{Solution of loop  equation in $1/N^2$ approximation}

The main goal of this paper is to find function $F^{1}$ on our moduli space which satisfies the equation
\be
\f{\de F^{1}}{\de\U(x)}=-Y^{1}(x)\;,
\la{eqf1}
\ee
where $Y^{1}(x)$ should be determined from $1/N^2$ expansion of the  loop equation (\ref{master}).
The equation (\ref{eqf1}) is valid in a neighbourhood of the point $\ix$ i.e in a neighbourhood of the point $x=\infty$ on the ``physical"
(with respect to variable $x$) sheet of the spectral curve $\L$. 
The same function $F^{(1)}$  should satisfy the equation
\be
\f{\de F^{1}}{\de\V(y)}=-X^{1}(y)\;,
\la{eqf1X}
\ee
where function $X^{1}(y)$ should be found from writing down the  loop equation with respect to
matrix $M_2$ in a neighbourhood of point $\iy$. We shall first solve (\ref{eqf1}), and then check the symmetry of the expression with respect
to the change of projection $\x\leftrightarrow \y$.

To express $Y^1$ in terms of the objects associated to the spectral curve $\L$ we consider the $1/N^2$ term of the
master loop equation (\ref{master}); we have:
\be
\Ecal(x,Y(x))= \Ecal^0(\x(P),\y(P)+\f{1}{N^2}Y^1(P)+\dots)+\f{1}{N^2}\Ecal^1 (\x(P),\y(P))+\dots\;,
\ee
as $P\to\ix$,
where, as before, in a neighbourhood of $\ix$, $\x(P)=x$; $\y(P)=Y^0(x)$. The $1/N^2$ expansion of $\Ecal(x,y)$ looks as follows:
\be
\Ecal(x,y)= \Ecal^{0}(x,y)+\f{1}{N^2}\Ecal^{1}(x,y)+\dots\;;
\la{Enext}
\ee
since $\Ecal^1(x,y)=-\Pcal^1(x,y)$, we can further rewrite this expression 
in a neighbourhood of point $\ix$ as follows:
\be
\Ecal(x,Y(x))= \Ecal^0(\x(P),\y(P)+\f{1}{N^2}\{ \Ecal^1 (\x(P),\y(P))+ Y^1(P)\Ecal^0_{y} (\x(P),\y(P))\}+\dots\;.
\ee
Therefore, the $1/N^2$ term of master loop equation (\ref{master}) gives
$$
\Ucal^0(\x(P),\y(P),\x(P))=\Ecal^1(\x(P),\y(P))+Y^1(P)\Ecal^0_y(\x(P),\y(P))
$$
as $P\to\ix$, or
\be
Y^1(P)=\f{\Ucal^0(\x(P),\y(P),\x(P))+\Pcal^1(\x(P),\y(P))}{\Ecal^0_y(\x(P),\y(P))}\;.
\la{Y1U0}
\ee
To make this formula more explicit we need to express 
  $\Ucal^0(\x(P),\y(P),\x(P))$ in terms of known objects using the  loop equation (\ref{loop1}).
According to definition of $\Ucal^0(x,y,z)$ we have:
\be
\Ucal^0(x,y,z)=-\f{\de \Ucal^0(x,y)}{\de \U(z)}
\ee
On the other hand, the zeroth order term of (\ref{loop1}) gives:
\be
\Ucal^0(x,y)=x-\V'(y)+\f{\Ecal^0(x,y)}{y-\y(x^{(1)})}
\ee
(as before,  $x^{(1)}$ denotes a point on the physical sheet of $\L$). 
Therefore,
\be
\Ucal^0(x,y,z)=-\f{\de\Ecal^0(x,y)/\de\U(z)}{y-\y(x^{(1)})}
-\f{\Ecal^0(x,y)}{(y-\y(x^{(1)}))^2}\f{\de \y(x^{(1)})}{\de\U(z)}\;.
\la{U0xyz}
\ee
Using the form (\ref{Ezeros}) of the polynomial $\Ecal^0(x,y)$, we can further rewrite this expression
as follows:
\be
\f{\de \Ecal^0(x,y)}{\de\U(z)}=-\Ecal^0(x,y)\sum_{k=1}^{d_2+1}\f{\de \y(x^{(k)})}{\de\U(z)}\f{1}{y-\y(x^{(k)})}\;.
\la{pEpU}
\ee 
Substituting this expression into (\ref{Ezeros}), we get
\be
\Ucal^0(x,y,z)=\f{\Ecal^0(x,y)}{y-\y(x^{(1)})}\sum_{k=2}^{d_2+1}\f{\de \y(x^{(k)})}{\de\U(z)}\f{1}{y-\y(x^{(k)})}\;.
\ee 
 Substituting $z=x=\x(P)$ and taking the limit $y\to \y(x^{(1)})$, we have:
\be
\Ucal^0(\x(P),\y(P),\x(P))=\Ecal^0_y(\x(P),\y(P))\sum_{k=2}^{d_2+1}\f{\de \y(x^{(k)})}{\de \U(\x(P))}
\f{1}{\y(P)-\y(x^{(k)})}
\ee
as $P\equiv x^{(1)}\to\ix$.
 Now (\ref{Y1U0}) can be rewritten as follows:
\be
Y^1(P)=\f{\Pcal^1(\x(P),\y(P))}{\Ecal^0_y(\x(P),\y(P))}+\sum_{Q\neq P\;:\;\x(Q)=\x(P)}\f{\de \y(Q)}{\de\U(\x(P))}
\f{1}{\y(P)-\y(Q)}
\la{Y1PE}
\ee
as $P\to\ix$, which can be further transformed, using the formula (\ref{Bx}) for the  
Bergmann bidifferential:
\be
Y^1(P)d\x(P)=\f{\Pcal^1(\x(P),\y(P))}{\Ecal^0_y(\x(P),\y(P))}d\x(P)
+\sum_{Q\neq P\;:\;\x(Q)=\x(P)}\f{B(P,Q)}{d\x(Q)}\f{1}{\y(P)-\y(Q)}\;;
\la{Y111}
\ee
now we see that the $1$-form $Y^1(P)d\x(P)$ can be analytically continued from a neighbourhood of $\ix$ 
to the whole $\L$.

\begin{lemma}\la{lemnonsing}
\vskip0.5cm
Let the spectral curve $\L$ (\ref{alcur}) be non-singular. Then
the 1-form $Y^1(P)d\x(P)$(\ref{Y111}) is a meromorphic 1-form  on the spectral curve $\L$ with poles 
up to fourth order only at the branch points $P_k$ i.e. at
  the zeros of differential $d\x(P)$.
\end{lemma}
{\it Proof.} 
Let us verify the non-singularity of the first term,
\be
\f{\Pcal^1(\x(P),\y(P))}{\Ecal^0_y(\x(P),\y(P))}d\x(P)\;,
\la{fterm}
\ee
of (\ref{Y111}) everywhere on $\L$. For finite $\x(P)$ the 1-form (\ref{fterm}) can be singular only 
at the zeros of 
$\Ecal^0_y(\x(P),\y(P))$, which, if the curve $\L$ is non-singular,  are by definition the branch points $P_k$; these zeros are of the 1st order and are
canceled by the zeros of $d\x(P)$ at the branch points. 

To study behaviour of (\ref{fterm}) at $\ix$ and $\iy$ we mention that the polynomial $\Pcal(x,y)$ (\ref{Pxy})
(and, therefore, also its first correction $\Pcal^1(x,y)$)
is of degree $d_1-1$ with respect to $x$ and $d_2-1$ with respect to $y$. However, we can say a bit more about 
$\Pcal^1(x,y)$. Namely, the coefficient of $\Pcal(x,y)$ in front of $x^{d_1-1}y^{d_2-1}$ equals 
$u_{d_1+1} v_{d_2+1}$, which does not have any higher corrections.
Therefore, the coefficient of  $\Pcal^1(x,y)$ in front of $x^{d_1-1}y^{d_2-1}$ vanishes. 

Now consider the behaviour of the 1-form (\ref{fterm}) near $\ix$. 
We have
$$
\Ecal^0_y(\x(P),\y(P))=-(\V'(\y(P))-\x(P))-(\U'(\x(P))-\y(P))\V''(\y(P))-\Pcal^0_y(\x(P),\y(P))\;,
$$
which has pole of order $d_1 d_2$ near $\ix$ as corollary of asymptotics (\ref{assix}) of function $\y(P)$ near
$\ix$. The 1-form $d\x(P)$ has at $\ix$ the pole of second order. The most singular contribution by
$\Pcal^1(\x(P),\y(P))$ at $\ix$ comes from the term $\x^{d_1-2}(P)\y^{d_2-1}(P)$; it has the pole of order
$d_1-2+d_1(d_2-1)=d_1 d_2-2$. Summing up, we see that  (\ref{fterm}) is non-singular near $\ix$.

Consider the 1-form (\ref{fterm}) near $\iy$. At this point $d\x(P)$ has pole of order $d_2+1$; the main contribution to $\Ecal^0_y(\x(P),\y(P))$ is given by the term $(\U'(\x(P))-\y(P))\V''(\y(P))$, which has pole
of order $d_1 d_2+d_2-1$. Finally, the main contribution to $\Pcal^1(\x(P),\y(P))$ comes from the term
$\y^{d_2-2}(P)\x^{d_1-1}(P)$, which has pole of order $d_1 d_2-2$. In total (\ref{fterm}) is non-singular at $\iy$,
too.

Consider now the second term of (\ref{Y111}):
\be
\sum_{Q\neq P\;:\;\x(Q)=\x(P)}\f{B(P,Q)}{d\x(Q)}\;.
\f{1}{\y(P)-\y(Q)}
\la{sterm}\ee
The Bergmann bidifferential is singular (has second order poles) only at coinciding arguments, i.e. when $P$ coincides with one of the branch points $P_k$. The denominator $\y(P)-\y(Q)$ vanishes only if $P$ coincides with $Q$, (i.e. both of them coincide with one of the branch points $P_k$).
It is slightly more complicated to see that zeros of $d\x(Q)$ don't give any poles outside of $P_k$.
Obviously, $d\x(Q)$ is singular if $P\to P_k$ and $Q=P^*$,
where $P^*$ is another point such that $\x(P^*)=\x(P)$ and $P^*\to P_k$ as $P\to P_k$.
However, $d\x(Q)$ is also singular if $Q$ coincides with one of the branch points $P_k$, while $P$ remains
on some other sheet, and does not tend to $P_k$ as $Q\to P_k$. In this case in the sum (\ref{sterm}) we have
two singular terms (with poles of first order), which correspond to $Q$ and $Q^*$; however, the residues
of these terms just differ by sign, and, therefore, the total sum (\ref{sterm}) remains finite outside
the branch points $P_k$ and infinities $\ix$ and $\iy$.

As $P\to \ix$, all points $Q$ in (\ref{sterm}) tend to $\iy$; thus all $d\x(Q)$ have pole of order $d_2+2$;
all other terms remain non-singular and non-vanishing. Therefore, (\ref{sterm}) has zero of order $d_2+1$
at $\ix$.

As $P\to \iy$, the situation is slightly more complicated.
If we enumerate the sheets of $\L$ such that, as $x\to \infty$, $x^{(1)}\to\ix$, and 
$x^{(2)},\dots,x^{(d_2+1)}\to\iy$, and assume that $P= x^{(d_2+1)}$, then (\ref{sterm}) can be split as follows:
\be
\f{B(x^{(1)},x^{(d_2+1)})}{d\x(x^{(1)})}\f{1}{\y(x^{(d_2+1)})-\y(x^{(1)})}+
\sum_{j=2}^{d_2}
\f{B(x^{(j)},x^{(d_2+1)})}{d\x(x^{(j)})}\f{1}{\y(x^{(d_2+1)})-\y(x^{(j)})}\;.
\la{ssterm}
\ee
As $x\to\infty$, the first term in (\ref{ssterm}) has zero of order two ($d\x(x^{(1)})$ has pole of order two, other multipliers remain non-singular and non-vanishing). In each term of the sum in (\ref{ssterm}) the 
Bergmann bidifferential has pole of second order as $x\to\infty$; however, $d\x(x^{(j)})$ has pole of order $d_2+1$,
and $\y(x^{(d_2+1)})-\y(x^{(j)})$ has simple pole as $x\to\infty$; therefore, the whole expression (\ref{ssterm})
is non-singular (and even vanishing) as $x\to \infty$.

\vskip0.5cm

\begin{remark}\rm
The condition of non-singularity of the spectral curve (\ref{alcur}) made in lemma \ref{lemnonsing}
means in physical language that the spectral curve has maximal possible
genus equal to $d_1 d_2-1$
for given 
degrees of polynomials $V_1$ and $V_2$. If the genus of the spectral curve
 is less than the maximal genus, it must be singular; then the  non-singularity of 
1-form $Y^1(P)df(P)$ at the double points can not be verified rigorously. However, this 
non-singularity is suggested by physical  consideration: since we assume that at the double points 
one does not have any eigenvalues of $M_1$ or $M_2$ in large $N$ limit 
(i.e. corresponding filling fractions are equal to zero), there is no physical reason for
corresponding resolvents to be singular at these points in large $N$ limit.
Therefore, in the sequel we shall assume that  $Y^1(P)df(P)$ is non-singular outside of branch points
of $\L$ both for maximal and non-maximal genus. We should mention that this assumption was also made
(explicitly or implicitly) in the previous papers \cite{Chekhov1,Akemann,Chekhov,Eynard2}.
\end{remark}

The singular parts of $Y^1(P)d\x(P)$ at the branch points $P_k$ can  be found from (\ref{Y111}). If, say, 
$P\to P_k$, the only term in (\ref{Y111}) which contributes to singular part at $P_k$  corresponds to $Q=P^*$.
Thus
 \be
Y^1(P)d\x(P)=\f{B(P,P^*)}{d\x(P^*)}
\f{1}{\y(P)-\y(P^*)}+ O(1)\;;\hskip1.0cm {\rm as} \hskip0.8cm P\to P_k\;.
\ee    
Consider the local expansion of all ingredients of this expression as $P\to P_k$ in terms of the local parameter 
$\lp_k(P)=\sqrt{\x(P)-\l_k}$:
$$
\y(P)=\y(P_k)+\lp_k(P) \y'(P_k)+\f{1}{2}\lp_k^2(P)\y''(P_k)+\f{1}{6}\lp_k^3(P)\y'''(P_k)+\dots\;,
$$
$$
\y(P^*)=\y(P_k)-\lp_k(P) \y'(P_k)+\f{1}{2}\lp_k^2(P)\y''(P_k)-\f{1}{6}\lp_k^3(P)\y'''(P_k)+\dots\;,
$$
$$
d\x(P^*)=2\lp_k(P)d\lp_k(P)\;,
$$
$$
B(P,P^*)=\left(\f{1}{(2\lp_k(P))^2}+\f{1}{6} S_B(P_k)+\dots\right)d\lp_k(P)(-d\lp_k(P))\;.
$$
Therefore,
$$
\f{1}{\y(P)-\y(P^*)}=
\f{1}{2\lp_k(P)\y'(P_k)}\left(1-\f{\lp_k(P)^2}{6}\f{\y'''(P_k)}{\y'(P_k)}\right)+\dots
$$
and, as $P\to P_k$,
\be
\f{B(P,P^*)}{d\x(P^*)}\f{1}{\y(P)-\y(P^*)}=\left\{-\f{1}{16\lp^4_k(P)\y'(P_k)} 
+\left(\f{1}{96}\f{\y'''(P_k)}{\y'^2(P_k)} -\f{S_B}{24\y'(P_k)}\right)\f{1}{\lp_k^2(P)}+O(1)\right\}{d\lp_k(P)}\;.
\la{expa}
\ee

Since, according to our assumption, the $1$-form $Y^1(P)d\x(P)$ is non-singular on $\L$ outside of the branch points,
we can express this 1-form in terms of differentials $B(P,P_k)$ and $D(P,P_k)$  (\ref{BDkdef}) using their behaviour
near $P_k$; 
\be
\Y1(P)d\x(P) =\sum_{k=1}^{m_1}\left\{-\f{1}{96\y'(\xp_k)}D(P,\xp_k) + \left[\f{\y'''(\xp_k)}{96\y'^2(\xp_k)} -\f{S_B(\xp_k)}{24\y'(\xp_k)}\right] B(P,\xp_k)\right\}\;;
\la{Y1BD}
\ee
as a result we rewrite the equation (\ref{eqf1}) for $F^1$ as follows:
\be
\f{\de F^{1}}{\de\U(\x(P))}d\x(P)=\sum_{k=1}^{m_1}\left\{\f{1}{96\y'(\xp_k)}D(P,\xp_k) + \left[-\f{\y'''(\xp_k)}{96\y'^2(\xp_k)} +\f{S_B(\xp_k)}{24\y'(\xp_k)}\right] B(P,\xp_k)\right\}\;.
\la{F1BD}
\ee

\begin{proposition}\la{propp}
A general solutions $F^1$ of the system (\ref{F1BD}) can be written as follows:
\be
F^1=\f{1}{2}\log\tau_\x+\f{1}{24}\log\left\{\prod_{k=1}^{m_1} \y'(P_k) \right\}+ C(\{v_k\},\{\e_\a\})
\la{F1prel}
\ee
where $C(\{v_k\},\{\e_\a\})$ is a function on our moduli space
depending only on coefficients of polynomial $\V$ and filling fractions $\{\e_\a\}$; function $\tau_\x$ 
(the Bergmann tau-function on Hurwitz space) is defined
by the system of equations with respect to the branch points $\{\l_k\}$:
\be
\f{\p}{\p \l_k}\log\tau_{\x} = -\f{1}{12} S_B(P_k)\;;
\la{berg1}
\ee
function $\tau_\x$ depends on coordinates $\{u_k,\;v_k,\;\e_\a\}$ as a composite function.
\end{proposition}

{\it Proof.} The derivative of $\tau_f$ with respect to $\U(\x(P))$ is computed by  chain rule using
variational formula (\ref{zerothor}); derivatives of $\y'(P_k)$ with respect to $\U(\x(P))$ are given
by (\ref{yV1}). Collecting  all these terms together we see that derivative of (\ref{F1prel}) coincides with
(\ref{F1BD}).
\vskip0.5cm

Therefore, to  compute $F^1$ it remains to find the Bergmann tau-function $\tau_\x$ and to make sure that 
``constant" $C(\{v_k\},\{\e_\a\})$ is chosen such that the final expression is symmetric with respect to
the change of ``projection" i.e that $F^1$ satisfies also equations (\ref{eqf1X}).

\section{$F^{(1)}$ and Bergmann tau-function on Hurwitz spaces}

\subsection{Bergmann tau-function on Hurwitz spaces}

Here, following \cite{Duke}, we discuss the Bergmann tau-function on Hurwitz spaces for the stratum of Hurwitz space
arising in the application to two-matrix model.

The Hurwitz space $H_{g,N}$ is the space of equivalence classes of  pairs ($\L$, $\pib$), where $\L$ is a compact Riemann surface of genus $g$ and $\pib$ is a 
meromorphic functions of degree $N$. 
The Hurwitz space is stratified  according to multiplicities of poles 
of function $\pib$.
By $H_{g,N}(k_1,\dots,k_n)$, where $k_1+\dots+k_n=N$, we denote the stratum of $H_{g,N}$ consisting of meromorphic functions  
which have $n$ poles on $\L$ with multiplicities $k_1,\dots,k_n$.
(In applications to two-matrix model we need to study the tau-function on the stratum $H_{g,N}(1,N-1)$, on which the function
$\pib$ has only two poles: one simple pole and one pole of order $N-1$.)
 
Suppose that all critical points of the function $\pib$ are simple; denote them by 
$P_1,\dots P_M$ ($m=2N+2g-2$ according to Riemann-Hurwitz formula); 
the critical values
$\l_k=\pi(P_k)$ can be used as (local) coordinates on $H_{g,N}(k_1,\dots,k_n)$. Function $\pib$ defines  the realization of the 
Riemann surface $\L$ as an $N$-sheeted branched covering 
of $\CP1$ with ramification points $P_1,\dots, P_m$ and branch points $\l_k=\pib (P_k)$; the points at infinity
we denote by  $\infty_1,\dots,\infty_n$. 
In a neighbourhood of the ramification point $P_k$ the  local coordinate
is chosen to be $x_k:=\sqrt{\l-\l_k}$, $k=1,\dots,m$; in a neighbourhood of the point $\infty_j$ the 
local parameter is $x_{m+j}:=\l^{-1/k_j}$

 The Bergmann bidifferential $B(P,Q)$ has the second order pole as $Q\to P$ with the 
local behaviour (\ref{Bdia}): $B(P,Q)/\{d\lp(P) d\lp(Q)\}=(\lp(P)-\lp(Q))^{-2} +\f{1}{6}S_B(\lp(P))+o(1)$, 
where $\lp(P)$ is a local coordinate; $S_B(\lp(P))$ is the Bergmann
projective connection. 

We define the  Bergmann $\tau$-function $\tau_{f}(\l_1,\dots,\l_m)$ locally  by the system of equations (\ref{berg1}):
\begin{equation} 
\f{\p}{\p \l_k}\log\tau_{\pib} = -\f{1}{12} S_B(\lp_k)|_{\lp_k=0}\;,\hskip0.8cm k=1,\dots,m\;.
\label{deftau}
\end{equation}
compatibility of this system is a simple corollary of Rauch variational formulas \cite{MPAG}.

Consider the divisor  of the differential $d\pib$:
$(d\pib)=\sum_{k=1}^{m+n} r_k D_k$
where $D_k:=P_k\;,\;\; r_k:=1$ for $k=1,\dots,m$ and 
$D_{m+j}=\infty_{j}\;,\;\; r_{m+j}=-(k_j+1)$ for $j=1,\dots,n$; the corresponding local parameters $\lp_k$, $k=1,\dots,m+n$
were introduced above.

Here and below, if an argument of a differential coincides with a point $D_j$ of divisor $(df)$, we evaluate this
differential at this point with respect to local parameter $\lp_j$.
In particular, for the prime form we shall use the following conventions:
\begin{equation}
E(D_k,D_l):= {E(P,Q)}\sqrt{d\lp_k(P)}\sqrt{d\lp_l (Q)}|_{P=D_k,\;Q=D_l}\;,
\end{equation}
for $k,l=1,\dots,m+N$. 
The next notation correspond to prime-forms, evaluated at points of divisor $(df)$  with respect to  only one argument:
\begin{equation}
E(P,D_l):= {E(P,Q)}\sqrt{dx_l(Q)}|_{Q=D_l}\;,
\end{equation}
$l=1,\dots,m+n$;
in contrast to $E(D_k,D_l)$, which are just scalars, $E(P,D_l)$ are $-1/2$-forms with respect to $P$.

 Denote by $w_1,\dots,w_g$  normalized ($\oint_{a_\alpha} w_\beta=\delta_{\a\b}$ ) holomorphic differentials on $\L$;
$\B_{\a\b}=\oint_{b_\alpha} w_\beta$ is the corresponding matrix of $b$-periods; $\Th(z|\B)$ is the theta-function.

Choose some initial point 
$P\in\Lhat$ and introduce corresponding vector of Riemann constants $K^P$ and Abel map $\Acal_\a(Q)=\int_P^Q w_\a$.
Since zeros of differential $df$ have multiplicity 1, we can always choose the fundamental cell $\Lhat$
of the universal covering of $\L$  in such a way 
that $\Acal((df))=-2K^P$ (for an arbitrary choice of fundamental domain these two vectors coincide only up to
an integer combination of periods of holomorphic differentials), where the Abel map is computed along the path which does not intersect the boundary of $\Lhat$.

The key entry of the Bergmann tau-function is the following
holomorphic multivalued $(1-g)g/2$-differential ${{\cal C}}(P)$ on $\L$:
\be
{{\cal C}}(P):=\f{1}{W(P)}{\sum_{\a_1,\dots,\a_g=1}^g 
\frac{\partial^g\Theta(K^P)}{\p z_{\a_1}\dots \p z_{\a_g}} w_{\a_1}(P)\dots w_{\a_g}(P)}\;.
\ee
where
\be
W(P):= {\rm \det}_{1\leq \a, \b\leq g}||w_\b^{(\a-1)}(P)||
\la{Wronks}
\ee
denotes the Wronskian determinant of holomorphic differentials
 at point $P$.

The following theorem is a slight modification of the theorem proved in \cite{Duke}.

\begin{theorem}

The Bergmann tau-function (\ref{deftau}) of Hurwitz space  $H_{g,N}(k_1,\dots,k_n)$ is given by the following expression:

\begin{equation}
\tau_{\pib} = {{\cal Q}}^{2/3}  \prod_{k,l=1\;\;k< l}^{m+n} [E(D_k, D_l)]^{\frac{r_k r_l}{6}} \; 
\la{tauint}
\end{equation}
where the quantity ${{\cal Q}}$  defined by
\begin{equation}
{{\cal Q}} = [d\pib(P)]^{\f{g-1}{2}} {{\cal C}}(P)
\prod_{k=1}^{m+N}[ E (P,D_k)]^{\f{(1-g)r_k}{2}}\;;
\la{Fdef}
\end{equation}
is independent of the point $P\in \L$. 

 \end{theorem}

The proof of the theorem is very similar to  \cite{Duke}. The only technical difference is the 
appearance of   higher order poles of function $\pib$.

\subsection{Dependence of Bergmann tau-function on  the choice of the projection}

\begin{theorem}\la{tauftaug}
Let $\tau_{\x}$ and $\tau_{\y}$ be Bergmann tau-functions (\ref{tauint}) corresponding to 
divisors $(d\x)$ and $(d\y)$,
respectively.
Then
\be
\left(\f{\tau_{\x}}{\tau_{\y}}\right)^{12}=C\f{(u_{d_1+1})^{1-\f{1}{d_1}}}{(v_{d_2+1})^{1-\f{1}{d_2}}}
\f{\prod_{k}d\x(\yp_k)}{\prod_k d\y(\xp_k)}
\la{ttau}\ee
where 
\be
C=\f{d_1^{d_1+3}}{d_2^{d_2+3}}
\la{const}
\ee
 is a constant independent of moduli parameters.
\end{theorem}
{\it Proof.} 
As above, we assume that the fundamental cell $\Lhat$ is chosen in such a way that $\Acal((d\x))= \Acal((d\y))=-2 K^P$.
Denote divisors $(d\x)$ and $(d\y)$ as follows:
\be
(d\x)= \sum_{k=1}^{m_1}\xp_k -2\ix - (\dV+1)\iy  :=\sum_{k=1}^{m_1+2} r_k D_k\;,
\ee
\be
(d\y)= \sum_{k=1}^{m_2}\yp_k -2\iy - (\dU+1)\ix  :=\sum_{k=1}^{m_2+2} s_k G_k\;.
\ee
Since ${\rm deg} (d\x)={\rm deg} (d\y)=2g-2$, we have $\sum_{k=1}^{m_1+2} r_k= \sum_{k=1}^{m_2+2} s_k =2g-2$.
Then, according to the expression (\ref{Fdef}) for the Bergmann tau-function, we have
\be
(\tau_{\x})^{12}= {{\cal C}}^8(P)[d\x(P)]^{4g-4}\prod_{k,j=1}^{m_1+2} \{E(D_k, D_j)\}^{2 r_k r_j}
\prod_{k=1}^{m_1+2} \{E(P,D_k)\}^{r_k(4-4g)}\;,
\ee
where the values of prime-forms at the points of divisor $(d\x)$ are evaluated in the system of local parameters
defined by function $\x$ i.e. near $\xp_k$ the local parameter is $\lp_k=\sqrt{\x(P)-\l_k}$; near $\ix$ the local parameter
is $\lp_{m_1+1}=1/\x(P)$, and near $\iy$ it is $\lp_{m_1+2}=[\x(P)]^{-1/\dV}$.

Similarly, we have
\be
(\tau_{\y})^{12}= {{\cal C}}^8(P)[d\y(P)]^{4g-4}\prod_{k,j=1}^{m_2+2} \{E(G_k, G_j)\}^{2 s_k s_j}
\prod_{k=1}^{m_2+2} \{E(P,G_k)\}^{s_k(4-4g)}\;,
\ee
where the values of prime-forms at the points of divisor $(d\y)$ should be evaluated in the system of local parameters
defined by function $\y$ i.e. near $\yp_k$ the local parameter is $\lpy_k=\sqrt{\y(P)-\mu_k}$; near $\ix$ the local parameter
is $\lpy_{m_2+1}=1/\y(P)$, and near $\iy$ it is $\lpy_{m_2+2}=[\y(P)]^{-1/\dV}$.

Therefore, 
\be
\left(\f{\tau_{\x}}{\tau_{\y}}\right)^{12}=\frac{\prod_{k,j=1}^{m_1+2} \{E(D_k, D_j)\}^{2 r_k r_j}}
{\prod_{k,j=1}^{m_2+2} \{E(G_k, G_j)\}^{2 s_k s_j}}\left\{\f{d\x(P)}{d\y(P)}
\f{\prod_{k=1}^{m_2+2}\{E(P,G_k)\}^{s_k}}{\prod_{k=1}^{m_1+2}\{E(P,D_k)\}^{r_k}}\right\}^{4g-4}\;.
\la{rat}
\ee

Using independence of this expression of the choice of point $P$ we can split the $(4g-4)$th power
in this formula into the product over points of divisor $(d\x)+(d\y)$ (whose degree equals exactly
$4g-4$!). The subtlety which arises is that, evaluating the prime-forms and differentials $d\x$ and $d\y$
at the points $D_k$ and $G_k$ we fix the local parameters (these local parameters at the points of $(d\x)$
are defined via function $\x$, and at the points of $(d\y)$ via function $\y$ as explained above).
Since divisors $(d\x)$ and $(d\y)$ have common points ($\ix$ and $\iy$),
in a neighbourhood of each of these points we introduce two essentially different local parameters,
 and it is important to remember in each case
in which local parameter the prime-forms are computed.

Another subtlety is that, being considered as functions of $P$, different multipliers in (\ref{rat}) 
either vanish or become singular if $P\in (d\x)+(d\y)$; cancellation of these singularities should be 
accurately traced down.

Consider the first ``half'' of this expression, namely, the product over $P\in (d\x)$:
\be
\left\{\f{d\x(P)}{d\y(P)}
\f{\prod_{k=1}^{m_2+2}\{E(P,G_k)\}^{s_k}}{\prod_{k=1}^{m_1+2}\{E(P,D_k)\}^{r_k}}\right\}^{2g-2}=
\prod_{l=1}^{m_1+2}\lim_{P\to D_l}\left\{\f{d\x(P)}{d\y(P)}
\f{\prod_{k=1}^{m_2+2}\{E(P,G_k)\}^{s_k}}{\prod_{k=1}^{m_1+2}\{E(P,D_k)\}^{r_k}}\right\}^{r_l}
\ee
\be
= \prod_{k,l=1\,,\,k<l}^{m_1+2} \{E(D_l,D_k)\}^{-2 r_k r_l}
\prod_{k=1}^{m_1+2}\left\{\lim_{P\to D_k}\f{d\x(P)}{\{E(P,D_k)\}^{r_k}}\right\}^{r_k}
\prod_{l=1}^{m_1+2}\left\{\lim_{P\to D_l}\f{\prod_{k=1}^{m_2+2}\{E(P,G_k)\}^{s_k}}{d\y(P)}\right\}^{r_l}\;.
\la{calcul1}
\ee
The first product looks nice since it cancels out against the first product in the numerator of (\ref{rat}).
Let us evaluate other ingredients of this expression.
We have $D_k=\xp_k\;,\;\;r_k=1$ for $k=1,\dots, m_1$, $D_{m_1+1}=\ix\;,\;\;k_{m_1+1}=-2$,
$D_{m_1+2}=\iy\;,\;\;k_{m_1+2}=-(\dV+1)$. Therefore,
$$
\prod_{k=1}^{m_1+2}\left\{\lim_{P\to D_k}\f{d\x(P)}{\{E(P,D_k)\}^{r_k}}\right\}^{r_k}
$$
\be
=
\left\{\lim_{P\to D_{m_1+1} } \{d\x(P) E^2(P, D_{m_1+1})\} \right\}^{-2}
\left\{\lim_{P\to D_{m_1+2}} \{ d\x(P) E^{\dV+1}(P, D_{m_1+2})\} \right\}^{-\dV-1}
\prod_{k=1}^{m_1}\lim_{P\to \xp_k}\f{d\x(P)}{\{E(P,\xp_k)\}}\;,
\la{calcul} 
\ee
where we don't write $\ix$ and $\iy$ instead of $D_{m_1+1}$ and $D_{m_1+2}$, respectively, to remember that
we need to use the system of local parameters related to $\x(P)$.
The last term in (\ref{calcul}) product is the easiest one:
\be
\lim_{P\to \xp_k}\f{d\x(P)}{\{E(P,\xp_k)\}}=\lim_{\lp_k(P)\to 0}\f{2\lp_k}{\lp_k}=2\;.
\ee
In a similar way we evaluate the first term:
\be
\lim_{P\to D_{m_1+1}} \{d\x(P) E^2(P, D_{m_1+1})\}= -1\;,
\ee
and the second one:
\be
\lim_{P\to D_{m_1+2}} \{ d\x(P) E^{\dV+1}(P, D_{m_1+2})\}=  -d_2\;.
\ee
It remains to evaluate the third product in (\ref{calcul1}):
$$
\prod_{l=1}^{m_1+2}\left\{\lim_{P\to D_l}\f{\prod_{k=1}^{m_2+2}\{E(P,G_k)\}^{s_k}}{d\y(P)}\right\}^{r_l}
=\left(\prod_{l=1}^{m_1} \{d\y (\xp_l)\}^{-1}\right) 
\left(\prod_{{\rm all}\; k,l\; {\rm such\; that}\; D_l\neq G_k} \{E(D_l,G_k)\}^{r_l s_k}\right)
$$
\be
\times\left(\lim_{P\to D_{m_1+1}}{\{E(P,G_{m_2+2})\}^{d_1+1}}{d\y(P)}\right)^{2}
\left(\lim_{P\to D_{m_1+2}}{\{E(P,G_{m_2+1})\}^{2}}{d\y(P)}\right)^{d_2+1}\;.
\la{limits}
\ee
Consider the first limit in (\ref{limits}):
\begin{lemma}
\be
\lim_{P\to D_{m_1+1}}\left(\{E(P,G_{m_2+2})\}^{d_1+1}{d\y(P)}\right)^2= (d_1^2) (u_{d_1+1})^{1-\f{1}{d_1}}  
\la{lim1}
\ee
\end{lemma}
{\it Proof.}
Two different  local parameters at the point $\ix\equiv D_{m_1+1}\equiv G_{m_2+2}$ which we need to use are 
$\lp_{m_1+1}(P) = {\x^{-1}(P)}$ and $\lpy_{m_2+2}(P)=\y^{-1/d_1}(P)$. 
We  have 
\be
E(P,G_{m_2+2})=\f{(\lpy_{m_2+2}(P)+ \dots)}{d\sqrt{\lpy_{m_2+2}(P)}}= 
\sqrt{\f{d\lp_{m_1+1}}{d\lpy_{m_2+2}}(\ix)}\f{(\lpy_{m_2+2}(P)+\dots)}{\sqrt{d\lp_{m_1+1}(P)}}\;.
\ee
We have also $\y(P)= \lpy_{m_2+2}^{-d_1}$; thus
\be
dg(P)= -(d_1) (\lpy_{m_2+2})^{-d_1-1} \left(\f{d\lpy_{m_2+2}}{d\lp_{m_1+1}}(\ix)\right)     d\lp_{m_1+1}(P)\;.
\ee
Taking in (\ref{lim1}) the limit $P\to D_{m_1+1}$ we indicate that all
differentials in the bracket should be evaluated with respect to the
local parameter $\lp_{m_1+1}$. Therefore, in (\ref{lim1}) we ignore
all factors $d\lp_{m_1+1}(P)$ and
(\ref{lim1}) turns out to be equal to 
\be
(d_1^2)\left(\f{d\lpy_{m_2+2}}{d\lp_{m_1+1}}(\ix)\right)^{1-d_1}=  (d_1^2) (u_{d_1+1})^{1-\f{1}{d_1}}  \;,
\la{lim1a}
\ee
where we take into account that, as $P\to \ix$, $\y=u_{d_1+1}x^{d_1}+\dots$; thus
$({d\lpy_{m_2+2}}/{d\lp_{m_1+1}})(\ix)=(u_{d_1+1})^{-{1}/{d_1}}$.
\vskip0.5cm

Consider now the second limit in (\ref{limits}):
\begin{lemma}
\be
\lim_{P\to D_{m_1+2}}\{E(P,G_{m_2+1})\}^{2}{d\y(P)}=-1
\la{lim2}\ee
\end{lemma}
{\it Proof.}
In analogy to (\ref{lim1}) we have to evaluate  the prime-form and the differential $d\y$ with respect to 
 the local parameter related to function $\x$ i.e. with respect to $\lp_{m_1+2}(P)=(\x(P))^{-1/d_2}$, while the
local parameter arising from function $\y$ is $\lpy_{m_2+1}(P)=(\y(P))^{-1}$.
We have near $D_{m_1+2}$:
\be
E(P,G_{m_2+1})=\f{\lpy_{m_2+1}(P)+\dots}{\sqrt{\lpy_{m_2+1}(P)}}=
\sqrt{\f{ \lp_{m_1+2}}{d\lpy_{m_2+1}}(\iy)}\f{\lpy_{m_2+1}(P)+\dots}{\sqrt{d\lp_{m_1+2}(P)}}\;.
\ee
and
\be
d\y(P)=d\left(\f{1}{\lpy_{m_2+1}(P)}\right)=-\f{d\lpy_{m_2+1}}{ \lp_{m_1+2}}(\iy)
\f{d\lp_{m_1+2}(P)}{ \lpy^2_{m_2+1}(P)}\;.
\ee
As before, substituting these expressions to (\ref{lim2}) and ignoring the arising power of $d\lp_{m_1+2}(P)$,
we see that this limit equals $-1$.
\vskip0.5cm

Substituting this $-1$, together with the answer (\ref{lim1a}) for the limit (\ref{lim1}), into (\ref{limits}),
and collecting all terms in (\ref{calcul1}),
we get 
$$
\left\{\f{d\x(P)}{d\y(P)}
\f{\prod_{k=1}^{m_2+2}\{E(P,G_k)\}^{s_k}}{\prod_{k=1}^{m_1+2}\{E(P,D_k)\}^{r_k}}\right\}^{2g-2}
$$
\be
=\left\{2d_1^2 d_2^{-(d_2+1)}\right\} (u_{d_1+1})^{1-\f{1}{d_1}}
\left(\prod_{l=1}^{m_1} \{d\y (\xp_l)\}^{-1}\right) 
\f{\prod_{D_l\neq G_k} \{E(D_l,G_k)\}^{r_l s_k}}
{\prod_{k,l=1\,,\,k<l}^{m_1+2} \{E(D_l,D_k)\}^{2 r_k r_l}}\;.
\la{prod1}
\ee

Now, computing the second ``half" of (\ref{rat}) i.e. taking the product analogous to (\ref{calcul1})
over points of divisor $(d\y)$, and taking the product with (\ref{prod1}),  we get the statement of theorem
\ref{tauftaug}.

\subsection{Bergmann tau-function and $F^{(1)}$}

\begin{theorem}
The $F^1$ solution of equations (\ref{eqf1}), (\ref{eqf1X}) (\ref{Y1BD}) is given by any of the following two 
equivalent formulas:
\be
F^{1}=\f{1}{24}\log\left\{\tau^{12}_{\x} (v_{\dV+1})^{1-\f{1}{\dV}}\prod_{k=1}^{m_1} d\y(\xp_k) \right\}+
\f{d_2+3}{24}\log d_2+C
\la{F11}
\ee
or
\be
{F}^{(1)}=\f{1}{24}\log\left\{\tau_{\y}^{12}(u_{\dU+1})^{1-\f{1}{\dU}}\prod_{k=1}^{m_2} d\x(\yp_k)\right\}
+\f{d_1+3}{24}\log d_1+C\;.
\la{F12}
\ee
Here $\tau_{\x}$ and  $\tau_{\y}$ are Bergmann tau-function (\ref{tauint})
built from divisors $(d\x)$ and $(d\y)$, respectively; $C$ is a constant.
\end{theorem}

{\it Proof.} From the formulas (\ref{ttau}), (\ref{const}) it follows  that 
expressions (\ref{F11}) and (\ref{F12}) define the same function.
According to Proposition \ref{propp}, expression (\ref{F11}) satisfies equations (\ref{eqf1}), (\ref{Y1BD})
 with respect to coefficients of $\U$. Similarly, expression (\ref{F12}) satisfies analogous system 
(\ref{eqf1X}) with respect to coefficients of $\V$.


\begin{remark}\rm (higher order branch points)
If potentials $\U$ and $\V$ are non-generic i.e. some (or all) of the branch points have multiplicity higher 
than 1, formula (\ref{F12}) should be only slightly modified. Namely, the expression for Bergmann tau-function
(\ref{tauint}) formally remains the same in terms of divisor of differential $d\x$ (the zeros of $d\x$ can now have
arbitrary multiplicities). The expression for $F^1$ then looks as follows:
\be
F^{1}=\f{1}{48}\log\left\{\tau^{24}_{\x} (v_{\dV+1})^{2-\f{2}{\dV}}\prod_{k=1}^{m_1} 
\res|_{P_m}\f{(d\y)^2}{d\x} \right\}+
\f{d_2+3}{24}\log d_2+C\;.
\la{higherorder}
\ee
The proof of (\ref{higherorder}) is slightly more involved technically than the generic case and will
be published separately. 
\end{remark}

\section{Equations with respect to filling fractions}

it is well-known (see for example \cite{Marco2}) that the  normalized ($\oint_{a_\a}w_\b=\delta_{ab}$)  holomorphic differentials can be expressed as follows in terms
of $\x$ and $\y$:
\be
2\pi i w_\a(P) = \f{\p \y(P)}{\p \e_\a}\Big|_{\x(P)} d\x(P)
\la{yea1}
\ee
(Sketch of the proof: differentiating  (\ref{filfr}) with respect to $\e_\b$, we verify the normalization conditions
for differentials (\ref{yea1}).
The 1-form $\y d\x$  is singular at $\ix$ and $\iy$; at $\ix$ we have $\y=\U'(\x)-1/\x+...$; this singularity 
disappear since coefficients of $\U$ and $\V$
are independent of filling fractions. Singularities at branch points $\xp_k$ of derivative of $\y$ with respect to $\e_\a$
get cancelled by zeros of $d\x$ at these points.
At $\iy$ we have: $x=\V'(\y)-1/\y+...$; due to thermodynamic identity 
$$
\f{\p \y}{\p\e_\a}\Big|_\x d\x= -\f{\p \x}{\p\e_\a}\Big|_\y d\y\;.
$$
Since coefficients of $\V$ are independent of $\e_\a$, singularity of $\y d\x$ at $\iy$ also disappears after
differentiation.)

To obtain equations for derivatives of $F^{(1)}$ with respect to the filling fractions
we shall prove the following analog of lemma \ref{deryU}:
\begin{lemma}
The following deformation equations with respect to filling fractions take place:
\be
\p_{\e_\a}\l_k =-2\pi i \f{w_\a(\xp_k)}{\y'(\xp_k)}
\la{lea}
\ee
\be
\f{\p\{\y'(\xp_k)\}}{\p\e_\a}= \f{\pi i}{2}\left\{w_a''(\xp_k)-\f{\y'''(\xp_k)}{\y'(\xp_k)} w_\a(\xp_k)\right\}\;.
\la{yea}
\ee
\end{lemma}

{\it Proof} is parallel to the proof of (\ref{zerothor}) and (\ref{yV1}): 
from (\ref{yea1}) we have
\be
\f{\p \y(P)}{\p \e_\a}\Big|_{\lp_k(P)} d\x(P) - \f{\p \x(P)}{\p \e_\a}\Big|_{\lp_k(P)}d\y(P) =2\pi i w_{\a}(P)\;.
\la{inter1}
\ee
Substituting in (\ref{inter1}) the local expansions (\ref{yexp}) of $\y(P)$ and (\ref{dyexp}) of $d\y(P)$,
and expansion of $w_\a(P)$
\be
w_\a(P)=(w_\a(\xp_k) + w_\a'(\xp_k)\lp_k+\f{w_\a''(\xp_k)}{2}\lp_k^2+\dots)d\lp_k\;,
\ee
we get, since $\x(P)=\lp_k^2(P)+\l_k$ and $d\x(P)=2\lp_k(P)d\lp_k(P)$:
$$
(\p_{\e_\a}\y(\xp_k)+\lp_k  \p_{\e_\a}\y'(\xp_k) +\f{1}{2}\p_{\e_\a}\y''(\xp_k)+\dots)2\lp_k d\lp_k
-\p_{\e_\a}\x_k (\y'(\xp_k)+\y''(\xp_k)\lp_k +\f{1}{2}\y'''(\xp_k) \lp_k^2+\dots)d\lp_k
$$
$$
=2\pi i (w_\a(\xp_k) +w_\a'(\xp_k)\lp_k+\f{1}{2}w_\a''(\xp_k)\lp_k^2)d\lp_k\;.
$$
The zeroth order term gives (\ref{lea}).
Collecting the coefficients in front of $\lp_k^2$, and using(\ref{lea}), we get
(\ref{yea}).
\vskip0.5cm
\begin{theorem}
Derivatives of function $F^1$  (\ref{F11}), (\ref{F12}) with respect to the filling fractions look as follows:
\be
\f{\p F^1}{\p \e_\a}=-\oint_{b_\a}Y^1(P)d\x(P)\;,
\ee
where $Y^1d\x$ is defined by (\ref{Y1BD})\;.
\end{theorem}
{\it Proof.}  
The vectors of $b$-periods of these 1-forms $B(P,\xp_k)$ and $D(P,\xp_k)$ can be expressed in terms
of holomorphic differentials via the following standard formulas:
\be
\oint_{b_a} B(P,\xp_k)=2\pi i w_\a (\xp_k)\;,\hskip0.8cm 
\oint_{b_\a} D(P,\xp_k)=2\pi i w_\a''(\xp_k)\;.
\ee

Therefore, the
$b$-periods of the $1$-form $-\Y1(P)d\x(P)$ (\ref{Y1BD}) are given by the following expression:
\be
-\oint_{b_\a}\Y1(P)d\x(P) =2\pi i\sum_{k=1}^{m_1}\left\{-\f{w_a''(\xp_k)}{96\y'(\xp_k)}+\f{\y'''(\xp_k) 
w_a(\xp_k)}{96\y'^2(\xp_k)}+\f{S_B(\xp_k)w_a(\xp_k)}{24\y'(\xp_k)}\right\}\;.
\la{perY1}
\ee
On the other hand,
derivatives of $F^1$ (\ref{F11}) with respect to $\e_\a$ can be computed using 
(\ref{lea}), (\ref{yea}) and equations for Bergmann tau-function (\ref{berg1}), which also leads to (\ref{perY1}).

\section{$F^1$ of two-matrix model, isomonodromic tau-function, $G$-function of Frobenius manifolds,
and determinant of Laplace operator}

Here we outline some links between the expression (\ref{F11}), (\ref{F12}) for $F^1$ and 
other well-known objects.

\subsection{$F^1$, isomonodromic tau-function  and $G$-function of Frobenius manifolds}

We recall that the genus 1 correction to free energy in topological field theories is given by
so-called $G$-function of associated Frobenius manifolds. The $G$-function is a solution of
Getzler equation \cite{Getzler}; for Frobenius manifolds related to quantum cohomologies, the $G$-function
was intensively studied  as generating function of elliptic Gromov-Witten invariants
(see \cite{DZ,Manin} for references).
In \cite{DZ} it was found the following formula for $G$-function of an arbitrary $m$-dimensional 
Frobenius manifold:
\be
G=\log\f{\tau_I}{\prod_{k=1}^m \eta_{kk}^{1/48}}
\la{Gfun}
\ee
where $\tau_I$ is the Jimbo-Miwa tau-function of Riemann-Hilbert problem associated to a given Frobenius manifold 
\cite{Dub92}; $\eta_{kk}$ are elements of Egoroff-Darboux (pseudo) metric 
(written in canonical coordinates) corresponding to the Frobenius manifold.

One of the  well-studied classes of Frobenius manifolds arises from Hurwitz spaces \cite{Dub92}. For these Frobenius manifolds the 
isomonodromic tau-function $\tau_I$ \cite{Dub92} is related to Bergmann tau-function $\tau_{\x}$ (\ref{berg1})
as follows \cite{IMRN}:
\be
\tau_I=\tau^{-1/2}_{\x}\;,
\ee
where $\x$ stands for meromorphic function on Riemann surface $\L$. Therefore, the tau-function terms, which are the main ingredients of the formulas
(\ref{F11}) for $F^1$ and (\ref{Gfun}) for the $G$-function coincide (up to a sign, which is related to the 
choice of the sign in the exponent in the definition (\ref{part})  of the free energy).
The solution of Fuchsian system corresponding to tau-function $\tau_I$ is not known explicitly.
However, the same function $\tau_I$, being
multiplied with certain theta-functional factor, gives tau-function of an arbitrary Riemann-Hilbert problem with
quasi-permutation monodromy matrices which was solved in \cite{Koro00}.

The metric coefficients of Darboux-Egoroff metric corresponding to Hurwitz Frobenius manifold
are defined in terms of an ``admissible" 1-form $\varphi$, defining the Frobenius manifold:
\be
\eta_{kk}=\res|_{P_k}\f{\varphi^2}{d\x}\;.
\la{rotco}
\ee
If, trying to develop an analogy with our formula (\ref{F11}) for $F^1$, we formally choose $\phi(P)=d\y(P)$,
we get $\eta_{kk}=\y'^2(P_k)/2$ and the formula  (\ref{Gfun}) coincides with (\ref{F11}) up to 
small details like sign, additive constant, and the highest coefficient of polynomial $\V$
arising from requirement of symmetry $f\leftrightarrow g$.

Therefore, we got complete formal analogy between our expression (\ref{F11}) for $F^1$ and Dubrovin-Zhang formula 
(\ref{Gfun}) for $G$-function. Unfortunately, for the moment this analogy remains only formal, since, from the
point of view of Dubrovin's theory \cite{Dub92}, the differential $d\y$ is not admissible; therefore, the metric
$\eta_{kk}=\y'^2(P_k)/2$ 
built from this differential  is not flat, and, strictly speaking, it does not correspond to any Frobenius manifold.
Therefore, the true origin of the analogy between the $G$-function of Frobenius manifolds and $F^1$ 
still has to be explored.

\subsection{$F^1$ and determinant of Laplace operator}

Existence of close relationship between $F^1$ and determinant of certain Laplace operator was suggested
by several authors (see e.g. \cite{DV} for  hermitial one-matrix model, \cite{Eynard2}
for hermitian two-matrix model and, finally, \cite{ZabWie} for normal two-matrix model 
with simply-connected support of eigenvalues, where $F^1$ is claimed to coincide with determinant 
of Laplace operator in the domain with Dirichlet boundary conditions).

However, 
 in the context of hermitial two-matrix model 
(as well as in the case of hermitian one matrix model \cite{DV})
this relationship is more subtle. 

First, if we don't impose any reality conditions
 on coefficients of polynomials $\U$ and $\V$, function $F^1$ is holomorphic function of our  moduli parameters
(i.e. coefficients of $\U$, $\V$ and filling fractions), while $\det\Delta$ is always a real function.
The Laplace operator $\Delta^\x$ which should be playing a role here corresponds to the 
singular metric of infinite volume $|d\x|^2$.
 
This problem disappears if we start from more physical situation, when all these moduli parameters are real, 
as well as the branch points of the Riemann surface $\L$ with respect to both projections. In this case 
$F^1$ is real itself, as well as determinant of Laplace operator.
However, 
little is known about rigorous definition  for determinants of
such Laplace operators, although such objects were actively used by string theorists without
rigorous mathematical justification \cite{Knizhnik,Sonoda,DHoPho}.
According to empirical results of \cite{Sonoda}, the regularised
determinant of Laplace operator $\Delta^\x$ is given by the formula
\be
\f{\det \Delta^\x}{{\cal A}\det \Im \B} = C |\tau_\x|^2\;,
\la{condet}
\ee  
where ${{\cal A}}$ is a regularised area of $\L$, $\Delta^\x$ is Laplace operator defined in singular metric $|d\x(P)|^2$, $\B$ is the matrix of $b$-periods of $\L$, $C$ is a constant.

In the ``physical" case of real moduli parameters the empirical expression (\ref{condet}) for 
$\log\{\det \Delta^\x\}$ coincides with $F^1$ (\ref{F11})  up to a simple power and additional
 multipliers.

Therefore, the relationship between hermitial and normal two-matrix models \cite{ZabWie} on the level of $F^1$ is not as
straightforward as on the level of functions $F^0$ ($F^0$ for hermitian two-matrix mode can be obtained
from $F^1$ for normal two-matrix model by a   
simple analytical continuation \cite{Marco1,Marco2,Zabr}).

From the point of view of determinants of Laplace operators the theorem \ref{tauftaug} which tells how the
Bergmann tau-function depends on the projection choice is nothing but a version of Alvarez-Polyakov formula \cite{Alva},
which describes variation of $\det\Delta$ if  the metric changes within given conformal class.

\section{From two-matrix to one-matrix model: hyperelliptic curves}

Suppose that $d_2=1$, i.e. polynomial $\V$ is quadratic. Then integration with respect to $M_2$ in (\ref{part})
can be carried out explicitly, and we get
\be
Z_N\equiv e^{-N^2 F}=C\int dM e^{-N\tr V(M)}
\ee
where $M:=M_1$,  $V:=\U$ and $C$ is a constant. Hence, in this case (\ref{part})  gives rise to the partition function of one-matrix model.

For $d_2=1$ the function $\x(P)$ has two poles of  order $1$ at $\ix$ and $\iy$; thus, the spectral curve $\L$ is hyperelliptic and function $\x(P)$ defines  two-sheeted covering of $CP1$.
The number of branch points in this case  equals $m_1\equiv 2g+2$; as before, we call  them $\l_1,\dots,\l_{2g+2}$. 
The Bergmann tau-function (\ref{berg1}) for hyperelliptic curves was computed in \cite{KitKor};
in this case it admits the following, alternative to (\ref{tauint}), (\ref{Fdef}) expression:
\be
\tau_{\x} =\Delta^{1/4} \det {\bf A}
 \la{tauhyp}
\ee
where
\be
\Delta:=\prod_{j<k\,,\;j,k=1}^{2g+2}(\l_j-\l_k)\;,
\la{Wander}
\ee
 ${\bf A}$ is the matrix of $a$-periods of non-normalized holomorphic differentials on $\L$:
\be
{\bf A}_{\a\b}=\oint_{a_\a}\f{x^{\b-1}d x}{\nu}\;;
\ee
where
$$
\nu^2=\prod_{k=1}^{2g=2} (x-\l_k)
$$
is the equation of spectral curve $\L$.

Substituting formula (\ref{tauhyp}) into (\ref{F11}), and ignoring coefficient 
$v_{d_2+1}$ (it becomes part of constant $C$), we get the expression
\be
F^1=\f{1}{24}\log\left\{\Delta^3\,(\det{\bf A})^{12}  \prod_{k=1}^{2g+2} \y'(\l_k)\right\}
\la{F1hyp}
\ee
which agrees with previously known results
\cite{Akemann,KMT,Chekhov,DST}.

{\bf Acknowledgements} We thank M.Bertola, L.Chekhov, B.Dubrovin,
T.Grava, V.Kazakov, I.Kostov, M.Staudacher and S.Theisen for important discussions.
The work of BE was partially supported by the EC ITH Network
HPRN-CT-1999-000161.
The work of DK was partially supported by  NSERC, NATEQ and Humboldt foundation. 
AK and DK thank Max-Planck Institute for Mathematics in Bonn for support and nice working conditions.
DK thanks also  SISSA and CEA  for support and hospitality. BE thanks
CRM for support and hospitality.


\begin{thebibliography}{99}



\bibitem{Mehta} M.L.Mehta, Random matrices, 2nd edition, Academic
Press, NY (2001)

\bibitem{BleIts} P.M.Bleher and A.R.Its, eds., ``Random Matrix models
and their applications'', MSRI Research publications {\bf 40},
Cambridge Univ.Press (Cambridge, 2000)

\bibitem{DGZ} P. Di Francesco,  P.Ginzparg, J.Zinn-Zustin, ``2D
Gravity and Random matrices'', Phys.Rep.  {\bf 254}, 1 (1995)

\bibitem{GMW} T.Guhr, A.Mueller-Groeling, H.A.Weidenmuller, ``Random
Matrix theories in quantum physics': Common concepts'', Phys.Rep. {\bf
299}, 189 (1998)


\bibitem{Staudacher} M.Staudacher, Combinatorial Solution of the
Two-Matrix Model, Phys. Lett. {\bf B305}  332-338  (1993)
  
\bibitem{KazakovIsing}Daul, J.-M.; Kazakov, V. A.; Kostov, I. K.,
 Rational theories of $2$d gravity from the two-matrix
model, Nuclear Phys. {\bf B409}  311-338 (1993)

\bibitem{Rauch}
Rauch, H. E., A transcendental view of the space of algebraic Riemann
surfaces. Bull. Amer. Math. Soc., {\bf 71} (1965), 1-39




\bibitem{Chekhov1}J. Ambjorn, L. Chekhov, C.F. Kristjansen, Yu. Makeenko,
Matrix Model Calculations beyond the Spherical Limit
 Nucl.Phys. {\bf B404} (1993) 127-172; Erratum: {\bf B449} (1995) 681

\bibitem{Akemann}G. Akemann,
Higher genus correlators for the hermitian matrix model with multiple cuts,
Nucl.Phys. {\bf B482} (1996) 403-430

\bibitem{KMT}A.Klemm, M.Marino, S.Theisen,
 Gravitational corrections in supersymmetric gauge theory and matrix
models, JHEP 0303 (2003) 051

\bibitem{DST}  R.Dijkgraaf, A.Sinkovics, M.Temurhan,
Matrix Models and Gravitational Corrections,
hep-th/0211241 

\bibitem{Kostov} I.Kostov, unpublished notes

\bibitem{Chekhov} L.Chekhov, Genus one correction to multi-cut matrix
model solutions, hep-th/0401089


\bibitem{Eynard1} B.Eynard, Large N expansion of the two-matrix model, hep-th/0210047

\bibitem{Eynard2} B.Eynard, Large N expansion of the two-matrix model, multi-cut case, math-ph/0307052

\bibitem{Marco1} M.Bertola, Free energy of the two-matrix model/dToda
tau-function, hep-th/0306184,
Nucl.Phys.B, to appear

\bibitem{Marco2} M.Bertola, Second and third order observables of the two-matrix model, hep-th/0309192

\bibitem{Dub92} B.Dubrovin, Geometry of 2d topological field theories, 
Integrable systems and quantum groups (Montecatini Terme, 1993),
120-348, Lecture Notes in Math. {\bf 1620},
Springer, Berlin, 1996.

\bibitem{KKprep}A.Kokotov, D.Korotkin, Determinants of Laplacians over
Riemann surfaces in flat metrics, in preparation


\bibitem{Duke} A.Kokotov, D.Korotkin, Bergmann tau-function on Hurwitz spaces 
and its applications, math-ph/0310008

\bibitem{MPAG} A.Kokotov, D.Korotkin, Tau-functions on Hurwitz spaces, math-ph/0202034
Mathematical Physics, Analysis and Geometry (2004), to appear


\bibitem{KazMar} V.Kazakov, A.Marshakov,
Complex Curve of the Two Matrix Model and its Tau-function,
J.Phys. {\bf A36} (2003) 3107-3136
           
\bibitem{KitKor} A.Kitaev, D.Korotkin, 
On solutions of the Schlesinger equations in terms of
$\Theta$-functions. IMRN {\bf 1998} No. 17, 877-905

\bibitem{DZ}B.Dubrovin, Y.Zhang, Bi-Hamiltonian hierarchies in $2$D
topological field 
theory at one-loop approximation, Comm. Math. Phys. {\bf 198} 311-361 (1998)

\bibitem{IMRN} A.Kokotov, D.Korotkin, On G-function of Frobenius
manifolds related to Hurwitz spaces, IMRN {\bf 2004} No.6, math-ph/0306053 

\bibitem{Koro00}D.Korotkin, Solution of matrix Riemann-Hilbert
problems with quasi-permutation monodromy matrices,
math-ph/0306061, Math. Annalen (2004), to appear

\bibitem{DV}R.Dijkgraaf, C.Vafa, On Geometry and Matrix Models, hep-th/0207106 

\bibitem{ZabWie} P.Wiegmann, A.Zabrodin,
 Large N expansion for normal and complex matrix ensembles,
hep-th/0309253

\bibitem{Dubr}Dubrovin B., Zhang Y., Bihamiltonian hierarchies in 2D topological field theory at
one-loop approximation, Commun. Math. Phys., 198 (1998), 311-361

\bibitem{Getzler}
Getzler, E. Intersection theory on $\overline{{cal M}}\sb {1,4}$ and
elliptic Gromov-Witten invariants. J. Amer. Math. Soc. {\bf  10}
973--998 (1997)

\bibitem{Manin} Manin Yu. I., Frobenius manifolds, quantum cohomology, and moduli spaces,
AMS, 1999


\bibitem{Knizhnik}Knizhnik, V.G. Multiloop amplitudes in the theory of
quantum strings and complex geometry,
 Soviet Phys. Uspekhi {\bf 32} No. 11, 945-971 (1990); 

\bibitem{DHoPho}E.D'Hoker, D.H.Phong, Functional determinants on
Mandelstam diagrams. 
Comm. Math. Phys. {\bf 124} 629-645 (1989)

\bibitem{Sonoda} H.Sonoda, Functional determinants on punctured
Riemann surfaces and 
their application to string theory, Nucl.Phys. {\bf B284} 157-192 (1987)

\bibitem{Zabr} A.Zabrodin,   Dispersionless limit of Hirota equations
in some problems of complex analysis, math.CV/0104169 

\bibitem{Alva}O.Alvarez,  Fermion determinants, chiral symmetry, and
the Wess-Zumino anomaly. 
Nuclear Phys. {\bf B238} 61-72 (1984)

\end{thebibliography}
\end{document}